\documentstyle[ichep,12pt]{article}
\begin{document}
\finalcopy
\vskip-2pc
{\bf \hfill HU-SEFT R 1993-12}
\vskip 2cm

\centerline{\bf FROM THE DEUTERON TO DEUSONS,}

\centerline{\bf  AN ANALYSIS OF DEUTERONLIKE MESON-MESON BOUND STATES}
\vskip 1cm
\centerline{Nils A. T\"ornqvist}
\centerline{Research Institute for High Energy Physics, SEFT}
\centerline{University of Helsinki}
\centerline{PB 9, FIN-00140 Helsinki 14, Finland}
\vskip 1.5cm

{\bf Abstract}

A systematic study of possible
deuteronlike two-meson bound states, {\it deusons}, is presented.
Previous arguments that many such bound states
may exist are elaborated with detailed arguments and numerical
calculations including,
in particular, the tensor potential. This tensor potential which is crucial for
the deuteron binding is shown to be very important also in the mesonic case.
Especially, in the  pseudoscalar $^3P_0$ pseudoscalar-vector ($P\bar V$) and
vector-vector ($V\bar V$) channels the important observation is made
that the centrifugal barrier from the P-wave  can be
overcome by the $1/r^2$ and $1/r^3$ terms of the tensor potential.

In the heavy meson  sector one-pion exchange alone
is strong enough to form at least deuteron-like
$B\bar B^*$ and  $B^*\bar B^*$ composites bound by approximately 50 MeV.
Composites of  $D\bar D^*$ and  $D^*\bar D^*$ states bound
by pion exchange alone are
expected near the thresholds, while in the light meson sector one generally
needs some additional short range attraction to form bound states.

The quantum numbers of these states are $I=0$, and
 $J^{PC}= 0^{-+},\ 1^{++}$ for the $P\bar V$ states and
$I=0$,  $J^{PC}=0^{++},\ 0^{-+},\ 1^{+-}$
and $2^{++}$ for the $V\bar V$ composites.
In $B\bar B^*$ the states: $\eta_b(\approx 10545),\
\chi_{b1}(\approx 10562)$ are predicted
and in $B^*\bar B^*$ one finds the states: $\eta_b(\approx 10590),\
\chi_{b0}(\approx 10582),\
h_b(\approx 10608),\ \chi_{b2}(\approx 10602)$.
Near the  $D\bar D^*$ threshold the states:
$\eta_c(\approx 3870),\ \chi_{c0}(\approx 3870)$ are predicted, and near the
$D^*\bar D^*$ threshold one finds the states: $\chi_{b0}(\approx 4015),\
\eta_{c}(\approx 4015),\  h_c(\approx 4015),\ \chi_{c2}(\approx 4015)$.

Within the light meson sector pion exchange gives strong attraction
for $P\bar V$ and $V\bar V$ systems with  quantum numbers where the  best
non-$q\bar q$ candidates exist, although pion exchange alone is not strong
enough to support such bound states. Thus, athough one cannot conclude
with certainty it to be the case, this fact  does favour
the picture  that
the the $\eta(1440)$ and the  $f_1(1420)$ are mainly $K\bar K^*$ composites
and the $f_0(1710)$ mainly  a $K^* \bar K^*$ bound state, while the
$f_0(1515)$ and $f_2(1520)$ could be predominantly
$(\rho\rho\pm\omega\omega)/\sqrt 2$ composites.
If the predicted $D\bar D^*$
and $D^*\bar D^*$ states are found, these would support this
interpretation of the light states.

In channels with exotic flavour or CP quantum numbers
pion exchange is generally repulsive or
quite weak. Therefore one does not expect
that such deuteronlike bound states exist, although
$B^* B^*$ may be an exception.

\eject
\onehead{\bf 1. INTRODUCTION}

The experimental fact (See Ref.\cite{PDG}) that there exists a handful of
mesonic
states, which certainly cannot find a place within the conventional
$q\bar q$ model is a challenging problem within particle physics. This
has motivated many papers on gluonium   or "molecular multiquark"
 models. The latter
can be divided into two kinds, either baglike multiquark models or
models for bound states of two or more hadrons.
In a recent paper\cite{NAT1} I suggested  that a subclass of
the latter, or {\it "deuteronlike meson-meson states"} where pion exchange
plays
a domiminant r$\hat{\rm o}$le, may turn out to be important, and may explain
the
seen candidates.  For such states I suggested the acronym {\it deuson},
(for reasons explained in more detail in Ref.\cite{NAT2}).
Needless to say, if the present non-$q\bar q$ candidates can be understood
as being deusons rather than gluonium states, it would be an important
experimental input to understanding QCD.

In this paper I shall elaborate this deuson model, and in particular study
the predictions for the heavy states containing two heavy mesons, where
the theoretical predictions turn out to be less ambiguous.
The question whether such bound states
actually can exist has been studied surprisingly sparsely
in the literature. The idea is of course not quite new, but has been
discussed generally only in passing whithin general
phenomenological models for meson-meson bound
states (See Refs.\cite{Okun}--\cite{ISGUR}),
where pion exchange
is not given special attention. Recently, after my first
letter\cite{NAT1} Ericson and Karl\cite{Karl} has also studied the
strength of pion exchange with similar conclusions,
 and Manohar and Weise\cite{Mano}  have studied especially
flavour exotic two $B^*$-meson bound states.

When trying to understand whether pion exchange can bind two hadrons
the deuteron is certainly the prime reference state, the existence of which
nobody  doubts. It has been studied in great detail over the years
(See Ref.\cite{Glen} and the recent reviews\cite{Rosa}, \cite {Torleif}).
There one knows that the dominant binding energy comes from
pion exchange between two colourless $qqq$ clusters - a proton and a
neutron.

I  shall use the same kind of approach, which has been so successful in
understanding the deuteron, generalized to meson-meson states.
This means that a rather simple nonrelativistic
potential model is developed and one finds
the bound states by solving the Schr\"odinger equation.
Of course,  this does not mean that sometimes a more sophisticated approach
with relativistic formalism, and a
more complicated dynamical theory etc. would be better. But, for  states
like the deuteron with small binding energy and comparatively heavy
constituents, a nonrelativistic treatment should already give a very good
first approximation, especially for heavy  constituents.
The simplicity of this approach also has another great
advantage: The results can
easily be  understood and checked by a broad class of readers.

In Sec.~2 the formalism for the universal pion exchange potential is summarized
when applied to $NN$ and meson-meson channels introducing the
relative coupling number, $\gamma$, as an overall strength parameter for the
potential.  In Sec.~3. the most important features, which can be
learnt from the deuteron are summarized, and Sec.~4 discusses how the numerical
solutions are obtained.
 Secs.~5 and 6 are the central sections of this paper where  the application
to pseudoscalar-vector ($P\bar V$) and, respectively,  the vector-vector
($V\bar V$) deusons are discussed, giving the quantum
numbers of states that should exist
or might exist.  Finally  Sec.~7 contains the conclusions.

\eject

\onehead{\bf 2. ONE-PION EXCHANGE}
The broken chiral symmetry of QCD predicts that the pion is singled out
as the by far lightest hadron. That the pion is very light, especially when
measured in terms of squared masses, has of course been experimentally
known for much longer time than QCD. Therefore, one has known for a long
time that the pion plays
a very important r\^ole in
the dynamics of hadrons, when long distance effects are important.
In nuclear physics pion exchange was traditionally
the first mechanism proposed for understanding nuclear binding. In fact,
the pion was predicted by Yukawa in 1938\cite{Yuk}
long before it was found experimentally in 1947\cite{Pion}.

The first step in studying the effects of virtual pions
is to find the one-pion exchange potential.
The modern way of deriving this is from the QCD Lagrangian using chiral
perturbation theory (See e.g. Ref.\cite{Mano}).
To make things simple, one can derive an effective
quark pion interaction
\begin{eqnarray}
{\cal L}_{int}= \frac g f \bar q(x) \gamma^\mu\gamma_5\vec
\tau q(x)\partial_\mu \pi(x) \ ,
\end{eqnarray}
\noindent
where $f$ is the pion decay constant $\approx 132$MeV
and  $g$ can be considered as an effective pion quark pseudovector coupling
constant. This
leads in a nonrelativistic approximation for heavy constituents, and
 using $SU6$ wave functions for the
hadrons, to the following effective Hamiltonians
for quarks, nucleons, vector mesons and for the pseudoscalar (P)
vector meson (V) transition
\begin{eqnarray}
H_{\pi qq}&=& -\frac g f (\vec \sigma\cdot \vec \nabla ) (\vec \tau\cdot
\vec \pi ) \label{Hqq} \ ,\\
H_{\pi NN}&=& -\frac {5g}{3f} (\vec \sigma\cdot \vec \nabla ) (\vec \tau\cdot
\vec \pi ) \label{HNN} \ ,\\
H_{\pi VV}&=& -\frac g f (\vec \Sigma \cdot \vec \nabla ) (\vec \tau\cdot
\vec \pi ) \label{HVV} \ ,\\
H_{\pi PV}&=& -\frac g f (\vec \epsilon\cdot \vec \nabla ) (\vec \tau\cdot
\vec \pi ) \label{HPV}\ ,
\end{eqnarray}
\noindent
where $\vec \sigma$ and $\vec \Sigma$ are
 the spin matrices for spin 1/2 and 1 respectively, $\vec \epsilon$
the polarisation vector for the vector meson and $\vec \tau$ the Pauli
isospin matrices.
The three first of these Eqs.~({\ref{Hqq}-\ref{HVV})
are operators to be sandwiched between spin 1/2 or spin 1
spinors, whereas for the $\pi PV$ transition the corresponding matrix element
is already taken. The factor $ 5/3$ in the pion nucleon coupling
of Eq.~(\ref{HNN}) is the
usual factor obtained in evaluating the matrix element for pion emission
from nucleons with $SU6\ qqq$ wave functions, which is perhaps more familiar
when it is similarily evaluated  for the axial vector coupling constant $g_A$.
These relations are not expected to
be exact, but are generally believed to
be right to within about 30\%, which is sufficient for our present
exploratory purpose.
\noindent

These relations lead to a one-pion exchange potential in momentum space
for I=1/2 constituents of the form
\begin{eqnarray}
V_{\pi}(NN) &=&
\frac {25}{9}\frac {g^2}{f^2}\ (\vec \tau_1\cdot\vec \tau_2)(\vec \sigma_1\cdot
\vec q~)
(\vec \sigma_2\cdot \vec q~) \ \frac 1 {\vec q^{~2}+m_\pi^2} \ , \label{VNN} \\
V_{\pi}(VV)=-V_{\pi}(V\bar V) &=& \frac {g^2}{f^2}
 (\vec \tau_1\cdot\vec \tau_2)(\vec \Sigma_1
\cdot \vec q~) (\vec \Sigma_2\cdot \vec q~) \ \frac 1
{ \vec q^{~2}+m_\pi^2}  \ ,\label{VVV} \\
V_{\pi}(PV\to VP) &=& \frac {g^2}{f^2}
(\vec \tau_1\cdot\vec \tau_2)(\vec \epsilon_1
\cdot \vec q~)(\vec \epsilon_2^{~*}\cdot \vec q~) \ \frac 1
{\vec q^{~2}+\mu^2}\ ,\label{VPV}
\end{eqnarray}
\noindent
where the constituents are assumed to be isodoublets,
and the normalization
 $\tau^2=\sigma^2=3,\ \Sigma^2=2$ is used. For $I=1,0$ constituents ($\rho,\
\omega$) only the isospin factor, $\vec \tau_1\cdot\vec \tau_2$,
needs to be replaced
by an $SU3_f$ weight (See Ref.\cite{NAT1}).
In the PV transition potential Eq.~(\ref{VPV})
$m_\pi^2$ is replaced by $\mu^2=m_\pi^2-(m_V-m_P)^2$, whereby one can take into
account, at least approximately, the recoil effect due to the fact
that the P and V constituents can have unequal masses.

Let us introduce the constant
\begin{equation}
V_0=\frac {m_\pi^3}{12\pi}\frac{g^2}{f^2}\approx 1.3 \ {\rm MeV}\ , \label{V0}
\end{equation}
the numerical value of which is fixed by the $\pi N$ coupling
constant ($f^2_{\pi N}/(4\pi ) =0.08=\frac {25} 9 \frac {g^2} {f^2} m_\pi^2$
from which $g\approx 0.6$ ).
This constant $V_0$  measures
the effective potential between  two quarks, when  $(S_{qq}=I_{qq}=1)$.
It is independent of the mass of an I=0 spectator quark. A good test of
this would be the width of the $D^*$ meson to $D\pi$. One predicts

\begin{equation}
\Gamma (D^*\to D^0\pi^+)= \frac 1{6\pi}\frac {g^2}{f^2}p^3_\pi
= 2V_0 \frac{p_\pi^3}{m_\pi^3} \ , \label{width}
\end{equation}

\noindent which gives 63.3 keV, and adding the $D^+\pi^0$ mode
using isospin one gets 92.6 keV for
the total hadronic width, well below the present experimental bound for the
total width, 131 keV\cite{ACCMOR}. The same formula gives for the
$K^*$ width 37 MeV, in reasonable agreement with the experimental 49.8 MeV.
(Normalizing to the $K^*$ width one would have $V_0=1.75$ MeV and
$g\approx 0.7$, from which the $D^*$ width would be 124 keV.)
One has here the natural situation, in accord with heavy quark symmetry,
that the strength of the pion cloud, which couples to
the light quarks only, does not depend on the heavy spectator quark mass.
Previously in contrast to this, one has often used
in analyses  for $V\to PP$ decays (e.g. Ref.\cite{Thews})
 a phenomenological Lagrangian  normalized in an  {\it ad hoc} way
with a coupling, $G_{VPP}/M_V$, because of dimensional reasons. This would
give a factor $(M_{K^*}/M_{D^*})^2\approx 1/5$
smaller $D^*$ width, when  one fixes the coupling by the $K^*$ width.
Hopefully the $D^*$ width will soon be measured,
which  will throw light on this old problem of which factors of mass
one should use when using flavour symmetry.

One can combine the formulas (\ref{VNN}-\ref{VPV}) to a single one by
collecting all the constants
into an overall number $\gamma$, the "relative coupling number".
In Ref.\cite{NAT1} I introduced a similar number "the relative binding number",
$RBN = -\gamma$. Because it is more natural
to have a positive number to measure
the strength of the binding, $\gamma$ is more convenient than $RBN$.
Thus for $NN$ one has $\gamma^{NN}_{SI}=
-\frac{25}{9}(\tau_1\cdot \tau_2)(\sigma_1\cdot\sigma_2)$,
for $(D^*\bar D)_\pm ,\ \gamma^{PV}_{I\pm}=\mp \tau_1\cdot \tau_2 $
and for $D^*\bar D^*\ \gamma^{VV}_{SI}=-(\tau_1\cdot
\tau_2)(\Sigma_1\cdot\Sigma_2)$. This number $\gamma$ measures
 the relative strength of the potential
compared to the contribution from one pair of
quarks in a spin triplet and isospin
triplet state for which $\gamma^{qq}_{SI}=-1$. For example for the deuteron
channel
 $\gamma^{NN}_{10}=25/3$ and for  $D^*\bar D^*$  in I=0, S=0,
$\gamma^{VV}_{00}=6$.
The larger this number $\gamma$ is, the stronger is the attraction,
and if it is negative there is repulsion. The universal one-pion exchange
potential in momentum space can then be written compactly:
\begin{eqnarray}
V_{\pi}(q) = -\gamma V_0 \frac {4\pi}{m_\pi}
 \ [D+S_{12}(q)]\frac 1 {\vec q^{~2}+m_\pi^2}\  ,
\end{eqnarray}
where $D$ is a diagonal matrix whose first element  is normalized to
unity. This is done by convenience in order to fix an overall
$\gamma$ in the case
when there are several channels. Often $D$ is simply the unit matrix.
Explicitely D and the
 tensor operator in momentum space $S_{12}$ are:
\begin{eqnarray}
D^{NN}&=&\vec \sigma_1\cdot \vec \sigma_2/
(\vec \sigma_1\cdot \vec \sigma_2)_{11} \ ,\\
D^{VV}&=&\vec \Sigma_1\cdot \vec \Sigma_2/
(\vec \Sigma_1\cdot \vec \Sigma_2)_{11} \ ,\\
D^{PV}&=&1 \ , \\
S_{12}^{NN}(q) &=& [3(\vec \sigma_1\cdot \vec q~) (\vec \sigma_2\cdot \vec q~)
/m_\pi^2
\ -\vec \sigma_1\cdot\vec \sigma_2]/<\vec \sigma_1\cdot \vec\sigma_2>_{11}\ ,
\\
S_{12}^{VV}(q) &=& [3(\vec \Sigma_1\cdot \vec q~) (\vec \Sigma_2\cdot \vec q~)
/m_\pi^2
-\vec \Sigma_1\cdot\vec \Sigma_2]/<\vec \Sigma_1\cdot \vec \Sigma_2>_{11}\ , \\
S_{12}^{PV}(q) &=& [3(\vec \epsilon_1\cdot \vec q~) (\vec \epsilon_2^{~*}
\cdot \vec q~)/m_\pi^2
\ -\ 1] \ .
\end{eqnarray}
\noindent Taking the Fourier transform one gets the potential
in configuration space
(omitting the delta function piece in the central potential, which
from the phenomenological point of view will be included in the short
range potential and regularization scheme discussed later). Thus the general
one-pion exchange potential as a function of $r$ is:
\begin{eqnarray}
V_{\pi}(r) = -\gamma V_0\ [D\cdot C(r)+S_{12}(\hat r)\cdot T(r)] \ ,
\end{eqnarray}
where $D$ is as above, $\hat r$ is the unit vector, and the $r$ dependence
is given by the functions
\begin{eqnarray}
C(r)&=& \frac {\mu^2}{m_\pi^2} \frac {e^{-\mu r}}{m_\pi r} \ ,\label{Cr} \\
T(r)&=& C(r) [1+\frac 3{\mu r} +\frac 3 {(\mu r)^2}] \ ,\label{Tr}
\end{eqnarray}
and the tensor operator in $r$ space:
\begin{eqnarray}
S_{12}^{NN}(\hat r)&=&
[3(\vec \sigma_1\cdot\hat r~)(\vec \sigma_2\cdot\hat r~)\ -
\vec \sigma_1\cdot\vec \sigma_2]/<\vec \sigma_1\cdot\vec \sigma_2>_{11}\ ,\\
S_{12}^{VV}(\hat r)&=&
[3(\vec \Sigma_1\cdot\hat r~)(\vec \Sigma_2\cdot \hat r~)-
\vec \Sigma_1 \cdot\vec \Sigma_2]/<\vec \Sigma_1\cdot\vec \Sigma_2>_{11}\ ,\\
S_{12}^{PV}(\hat r)&=&
[3(\vec \epsilon_1\cdot\hat r~)(\vec \epsilon^{~*}_2\cdot\hat r~)\ -
\ 1]\ .
\end{eqnarray}
The matrix elements of the tensor force, $S_{12}$
connect  different spin-orbit configurations and
are real numbers. E.g. for the deuteron $(^3S_1,\ ^3D_1)$ states these are
the well known numbers
$$ <\ ^3S_1|S_{12}|^3S_1>=0,\ <\ ^3S_1|S_{12}|^3D_1>=\sqrt 8 \ ,
<\ ^3S_1|S_{12}|^3D_1>= -2\ .$$
\noindent
For the different meson-meson configurations
similar matrix elements  will be given in the text.
Note that the  $r$ dependence in the tensor $T(r)$ contains singular $r^{-2}$
and $r^{-3}$
terms.  These make the tensor potential dominate at small $r$
(cf. Fig.~1) like an axial dipole-dipole interaction.

Because of this singular behaviour of the tensor potential it
 must be regularized at small distances. The perhaps most natural
 method is to introduce a form  factor at each $\pi N$ vertex, such as
$(\Lambda^2-\mu^2)/(\Lambda^2+t)$, which in $r$-space can be looked
upon as a spherical pion source with rms radius
\begin{equation}
 R=\frac{\sqrt {10}}{\Lambda} =\frac {0.624}{\Lambda / [{\rm GeV}]}\  {\rm fm}
 \ . \label{radius}
\end{equation}
This modifies the one-pion exchange potential  in $k$-space by this form
factor squared, i.e. by a "dipole form factor". In $r$-space the potential
is then modified such that instead of the forms of Eq.~(\ref{Cr}-\ref{Tr})
one has

\begin{eqnarray}
C_{\Lambda}(r)&=& \frac {\mu^2}{m_\pi^3 r}[e^{-\mu r}-e^{-\Lambda r}
- (\Lambda -\mu^2/\Lambda ) r  e^{-\Lambda r}]
 \ ,\label{Creg} \\
T_\Lambda (r)&=& e^{-\mu r}\frac {\mu^2     r^2+3\mu     r+3}{m_\pi^3 r^3} -
             e^{-\Lambda r}\frac {\Lambda^2 r^2+3\Lambda r+3}{m_\pi^3 r^3}
 - e^{-\Lambda r} \frac{(\Lambda^2-\mu^2)(\Lambda r +1)} {2 m_\pi^3 r} \  .
\label{Treg}
\end{eqnarray}

In Fig.~1  the potentials are shown, regularized this way with $\Lambda=1.2$
GeV. These forms are used in the numerical calculations presented here.
 The value of $\Lambda$ is rather poorly known phenomenologically.
In NN interactions  values between 0.8 and 1.5 GeV have been used depending on
the application, and whether $2\pi$ etc. exchanges are included or not.
Large values of $\Lambda$ are, however, favoured by scattering data
for NN phase shifts\cite{holinde}, which require $\Lambda >1.4$ GeV.
(See e.g. the discussion of Ericson and Rosa-Clot\cite{Rosa}.)
For the present applications to mesons and in particularily to heavy mesons,
which are known to be much smaller in size than nucleons,
one would expect a smaller $R$ of the pion source  in Eq.~(\ref{radius})
corresponding to a larger $\Lambda$. Again, a larger $\Lambda$  gives
 a stronger potential at very short distances increasing the binding
energy.
\vskip 1cm

\onehead{\bf 3. THE DEUTERON}
Our prime reference state is of course the deuteron, which is an isosinglet
$^3S_1-\ ^3D_1$ bound state of a proton and a neutron.
Since it is  a state, which
has been studied in great detail, for a long time,  one can
learn a lot from it. This will be
useful in understanding the mesonic counterparts of the deuteron.

The lowest spin parity NN states can be formed from the following
spin-orbitals:
\vskip 0.5cm

$\begin{array}{cccll}
J^P& S & I &{\rm Spin\ orbital}& \ \ \ \gamma_{SI}^{NN}\\
0^+& 0 & 0 &  ^1S_0           & +25/3 \\
0^-& 1 & 1 &  ^3P_0           & -25/9.\\
1^-& 0 & 1 &  ^1P_1           & -25   \\
1^+& 1 & 0 &  ^3S_1,\ ^3D_1,\ & +25/3,\ {\rm the\ deuteron}  \\
\end{array}$
\vskip 0.3cm

\noindent {\small Table 1. The quantum numbers of the lowest spin  states
of $NN$, and their relative coupling numbers $\gamma^{NN}_{SI}$.}

\vskip 0.5cm
\noindent
where in the last column the relative coupling numbers are listed,
$\gamma_{SI}^{NN}$,
defined in the previous section. For the deuteron
one defines conventional basis vectors $|^3S_1>$ and $|^3D_1>$
 such that the wave function is in general
$u(r)|^3S_1>+w(r)|^3D_1>$. The  deuteron potential $V_d(r)$ can then be written
in matrix form as:
\begin{eqnarray}
V_{d}(r) = -\frac{25}{3} V_0  \left[
\left( \begin{array}{cc} 1&   0      \\    0     & 1\end{array}\right) C(r) +
\left( \begin{array}{cc} 0& \sqrt 8 \\ \sqrt 8 & -2\end{array}\right) T(r)
\right]\ ,   \label{Vdeut}
\end{eqnarray}
where $r$ is the distance between the nuclei and $V_0,\ C(r).\ T(r)$ are
defined in the previous section
(Eqs.~\ref{V0},\ref{Cr},\ref{Tr} with $\mu=m_{\pi}$).

The overall strength of the potential
is given by $\gamma_{10}^{NN} V_0$, which expressed in terms of the
conventional pion nucleon coupling constant is
$ \gamma_{10}^{NN} V_0 = \frac {f^2_{\pi N}}{4\pi}\cdot f\approx 11.2$ MeV.

For the other three spin isospin NN channels listed above the potentials
$V_{SI}(r)$ are one dimensional:

\begin{eqnarray}
V_{00}(r)& =& -\frac{25}{3} V_0 C(r)
\ ,   \label{VNN00} \\
V_{01}(r)& =& +25 V_0 C(r)
\ ,   \label{VNN01} \\
V_{11}(r)& =& +\frac{25}{9} V_0 [ C(r)+2T(r)]
\ .   \label{VNN11}
\end{eqnarray}

The second  piece in Eqs.~(\ref{Vdeut}) and (\ref{VNN11})
comes from the tensor potential which for the deuteron
connects S and D waves, while in the single channel
$^3P_0$ potential (\ref{VNN11}) the tensor and the central terms simply add;
but being repulsive in the latter, it produces of course no $^3P_0$
bound state. In the
meson-meson analogue, there is a similar situation in the pseudoscalar
channels (c.f. Fig.~2), but with the
essential difference that the potential is attractive.

On the other hand, for the $^1S_0$  and $^1P_1$ states the tensor
potential is absent.
For the former, the $^1S_0$, the
relative coupling number is equally large as for
the deuteron $\gamma_{01}^{NN}= 25/3$, but the absence of the tensor piece
weakens the potential enough not to have a bound state.

There are a  few important
points to be learned from the deuteron and the $NN$ system,
which are essential also for the present application to meson-meson states:

(i) The central part of the one pion
potential $\gamma_{10}^{NN} V_0C(r)$ is insufficient to bind the
deuteron. By scaling arguments it is easy to show\cite{BW}
that a simple Yukawa potential can form a bound state only
if the well depth parameter satisfies the bound
\begin{equation}
 s= 0.5953\cdot (\gamma_{SI}^{NN} V_0)M/\mu^2>1 \ .\label{welldN}
\end{equation}
\noindent
 For the deuteron using the conventional pion
nucleon coupling constant this well depth parameter is
\begin{equation}
s_d=0.33\ , \label{welldd}
\end{equation}
\noindent
i.e., it falls well below the bound by a large factor of 3.
The nonexistence of a bound state in $|^1S_0>$ is a
manifestation of this fact, since there the central one pion
potential (\ref{VNN00}) is present
with the same strength as for the deuteron, but the tensor potential is
absent since  $<\ ^1S_0|S_{12}|^1S_0>=0$. However, the $^1S_0$ is almost
bound, only a few keV are missing, which implies that the remainder
from short range forces ($2\pi, "\sigma"$ or quark-gluon exchange)
gives  additional attraction, strong enough to compensate for the hard
core repulsion (from Fermi statistics of the 6 quark system, or $\omega$
etc. exchange).

This is of relevance also in the present study of
2-meson systems, where
one should expect the repulsive forces to be less significant,
at least in nonexotic channels,
and instead, the remaining short range forces to be even more
attractive. However, as will be discussed later,
a detailed reliable modelling of this is not an easy matter, especially
as the regularization of the tensor force also requires some
phenomenological input. But, as a rough measure of whether there
can be a bound state a well depth parameter of the same size of
(0.33) as for the deuteron, Eq.~(\ref{welldd})  is a reasonable signal.

(ii) The tensor potential
is very important in providing the binding\cite{Torleif}.
It is much larger in magnitude than
the central part at small distances, and in addition
 the off-diagonal matrix ele\-ment
$<\ ^3S_1|S_{12}|^3D_1>$ is large ($\sqrt 8$).
In Fig.~1  $C(r)$ and $T(r)$ and their regularized forms
$C_\Lambda (r)$ and $T_\Lambda (r)$ are plotted wit $\Lambda=1.2$ GeV,
and as can be seen $T(r)$  is much larger than $C(r)$
especially for small distances.
In fact, the tensor force will turn out to be
dominant also for all interesting  mesonic systems.

(iii) Although the potential is repulsive in the D wave, the off-diagonal
matrix element produces by iteration a bound state, which is
mainly in the S wave, but which has
a small D wave probability of approximately 6\%.
 This small D wave is, however, strong
enough to feed into the S wave and produce  the binding energy.
Thus the  fact that deuteron "has the freedom to
flip from the S wave to the D wave
and back" lowers the energy enough to make it bound, in spite of the fact
that it spends a comparatvely short time in the D wave, because of the
repulsive forces in that channel.
Thus although the tensor force appears in the deuteron
as a second order effect between the S
and the D wave, it is important for the binding. It will also sometimes be
even more important in the
case for deusons, where the tensor force can appear attractively in P wave
states as a first order effect.

(iv) The deuteron binding energy (2.22457 MeV) comes about as a small
net effect of comparatively large contributions of opposite signs. Its actual
value cannot be estimated by scaling arguments from the other scales involved
($M_N,\ m_\pi$, hadron size etc.).

(v) As  discussed in  Secs.~2 and 4 the
potential must be regularized at small $r$, by some {\it ad hoc}
cut off procedure. The binding energy is very sensitive to this procedure,
but once the binding energy is right the other deuteron properties
follow from the assumption of dominant pion exchange at larger distances.
To quote Glendenning and Kramer\cite{Glen} {\it "it happens to be one of the
quirks of nature that the force that binds the deuteron does it in such a
way as not to reveal itself too intimately"}.
For the same reason one cannot easily determine the contribution to the
short range potential coming from other exchanges than the pion
by studying deuteron properties alone.

This last point is unfortunate for us, since one will
not, in general, be able to predict reliably the binding energies for deusons.
Certainly, if the potential is strong enough one can be quite confident that
such bound states must exist, but their exact binding energy will always
depend on details of regularization or on the short range potential from
non-pion exchange. Again, in borderline
cases where the expected binding energy is small, like for the deuteron, one
similarily cannot be sure that  such bound states actually exist.
\vskip 2cm

\onehead{\bf 4. NUMERICAL METHOD}
The numerical solutions obtained in this paper are
 based on a method (See Refs.\cite{Basd}, \cite{Rich}
where one  discretizes the $r$ dependence to a finite dimensional vector,
wherby the Hamiltonian becomes a finite matrix.
Then the Schr\"odinger equation can be solved approximately as
a matrix equation, for which one can use efficient standard matrix routines
for finding the eigenvalues and eigenvectors. The method is particularily
accurate for finding the ground state,
which is precisely the problem of this paper.

This method is also particularily well suited when one  is
dealing with coupled channels, as is here the case. Namely,
for an n-channel problem the matrix dimension
of the  Hamiltonian simply becomes n times larger, in a rather trivial way.
E.g. for two channels as for the deuteron, 16 points on the $r$-axis
are sufficient to obtain  a very good approximation for the
binding energy and wave functions, by diagonalizing a 32 dimensional
matrix form of the Hamiltionian. The method is comphrehensively discussed
for the one-channel problem in Ref.\cite{Rich}.

In most of the numerical work presented here the regularization is done
by using the dipole form factor with a cut-off $\Lambda$ as discribed
by Eqs.~(\ref{Creg}-\ref{Treg}). But, I have also tried other methods of
regularization, e.g. the method suggested by Manohar and Wise\cite{Mano}
to simply saturate the potential
below a distance  $r_{cut} \approx 0.5$ fm to its value at
that distance: $V(r<r_{cut}) = V(r_{cut})$. Another method which have been
tried,  is to introduce a hard core\cite{Glen}
i.e., put $C(r)$ and $T(r)=\infty $ for some value
$r<r_{cut}< 0.5$ fm. Unfortunately there is some arbitrariness in
all such regularization procedures, but the dipole form factor,
which gives Eqs.~(\ref{Creg}-\ref{Treg}),
seems to be the most useful, both because
of its nice physical connection to the size of the pion source through
Eq.~(\ref{radius}), and because it seems to be the most widely used.

The deuteron S and D wave  functions obtained using this
numerical method, and using the dipole regularization are shown in Fig.~3,
and similarily for some of the mesonic systems in Figs.~4-6.
These agree of course very well with standard solutions found in the
literature, and with solutions obtained with much more detailed
$NN$ potentials (such as that of Ref.\cite{Paris}) apart from a small
region near $r=0$.
Here no other shorter range contributions than
one-pion exchange  are included,
 and the cut-off $\Lambda$ is fixed to get the right binding
energy. The deuteron quadrupole moment, D wave probability, rms radius and
other static properties are correctly predicted
close to their experimental values,
since these depend very little on the short range part of the potential,
once the binding energy is right.
\vskip 3cm

\onehead{\bf 5. PSEUDOSCALAR-VECTOR MESON STATES}

Since parity forbids two pseudoscalars to be bound by one-pion exchange,
the lightest deusons are pseudoscalar-vector (PV) states.
Again in  such $PV$ systems the  pion is too light to
be a constituent, because the small reduced mass of a $\pi V$ system would
give a too large kinetic term, which cannot be overcome by the potential.
In other words, the well depth parameter Eq.~(\ref{welldN}) becomes very small.
Thus the lightest deusons where pion exchange can play a dominant
r${\rm \hat o}$le are $K\bar K^*$ systems, i.e. the deuson spectrum can
start at $\approx 1400$ MeV.
\vskip 0.7cm

\noindent {\bf 5.1 Flavour nonexotic $(P\bar V)_{\pm}$ states.}
\vskip 0.4cm

Let us  denote the states with definite C parity $=\pm$ by
$(P\bar V)_\pm =[P\bar V \pm C(P\bar V)]/\sqrt 2$. Then
the states of lowest spin of one pseudoscalar and one vector meson,
with definite isospin and  C-parity
(for the neutral states) can appear in the following spin-orbital states:
\vskip 0.5cm

$
\begin{array}{lllll}
J^{PC}  &{\rm States}&\gamma_{0\pm}& \gamma_{1\pm}&<|S_{12}|> \cr
0^{-\pm}&   ^3P_0   &\pm 3&\mp 1&\ \ +2\cr
1^{-\pm}&   ^3P_1  & \pm 3 &\mp 1&\ \ -1\cr
1^{+\pm}&   ^3S_1,\ ^3D_1&\pm 3&\mp 1&{\rm See\ Eq.~(\ref{Vpvax})}
\end{array}
$
\vskip 0.3cm

\noindent
{\small Table 2. The lowest spin parity states for the $P\bar V$ systems,
the contributing partial waves,
their relative coupling numbers $\gamma_{IC}$, and $S_{12}$ matrix
elements. }
\vskip 0.5cm

 Table 2 also lists the values of the overall relative coupling number,
$\gamma_{IC}=C\cdot [2I(I+1)-3]$ and the relative strength of the tensor
potential in the single channel cases as measured
by the matrix element of $S_{12}$.
For the axial vector channel there
are two partial waves for the same spin-parity and these
mix, as is the case for the deuteron. The potentials in the three
cases are given below

\begin{eqnarray}
V_{0^{-\pm}}& =& -\gamma_{I\pm} V_0 [C(r)+2T(r)]=
-\gamma_{I\pm} V_0 \frac {\mu^2}{m_\pi^2}\frac{e^{-\mu r}}{m_\pi r} \left[ 3
 +\frac {6}{\mu r}+\frac {6}{(\mu r)^2} \right]\ , \label{Vpv0} \\
V_{1^{-\pm}}& =& -\gamma_{I\pm} V_0  [C(r)-T(r)]\ \ =
+\gamma_{I\pm} V_0 \frac{\mu^2}{m_\pi^2} \frac{e^{-\mu r}}{m_\pi r} \left[
 \frac {3}{\mu r}+\frac {3}{(\mu r)^2} \right]\ , \label{Vpv1} \\
V_{1^{+\pm}}& =& -\gamma_{I\pm} V_0  \left[
\left( \begin{array}{cc} 1&   0      \\    0     & 1\end{array}\right) C(r) +
\left( \begin{array}{cc} 0& -\sqrt 2 \\ -\sqrt 2 & 1\end{array}\right) T(r)
\right]\ ,  \label{Vpvax}
\end{eqnarray}

\noindent
where $V_0,\ C(r)$ and $T(r)$ are defined as in Eqs.~(\ref{V0}, \ref{Cr},
\ref{Tr}) and $\gamma_{I\pm}$ is given in Table 2.
Here $\mu$ is not exactly the pion mass, since the vector and
pseudoscalar do not have the same mass. This leads to a recoil effect
through which one has instead
\begin{equation}
\mu^2=m_\pi^2-(M_V-M_P)^2 \ .\label{mu}
\end{equation}

Physically this is a consequence of the fact that in the intermediate state
the pion can be closer to its mass shell.
For $K\bar K^*$ it can even be on the mass shell, and the simple
one pion exchange potential
should be supplemented with on-shell effects.
For $D\bar D^*$ $\mu$ almost vanishes and therefore, the
exponential Yukawa damping becomes less important. Because of the overall
$(\mu/m_\pi)^2$ factor the central potential (\ref{Cr})
is then less important and instead
the $r^{-3}$ dipole-dipole
part of the tensor potential (\ref{Tr}) becomes dominant.
For $B\bar B^*$ $\mu \approx 130$ MeV, i.e. it is
not very different from the pion mass and therefore, these effects are less
important there.

The pseudoscalar  channel is a single channel case, Eq.~(\ref{Vpv0}), to
which the tensor part contributes
with same sign as the central part since $<\ ^3P_0|S_{12}|^3P_0>=+2$.
Therefore, these add giving a remarkably strong potential (\ref{Vpv0}),
which is attractive for C=+ i.e., $J^{PC}=0^{-+}$.

For the case $1^{-\pm}$ there is also only one channel, but here
the central and tensor partly cancel since $<\ ^3P_1|S_{12}|^3S_1>=-1$
and one gets a somewhat weaker potential (\ref{Vpv1}) of opposite
overall sign containing
only  $r^{-2}$ and $r^{-3}$ terms from the tensor potential.
It would be most attractive for the  $1^{--}$ isoscalar channel.
In $K\bar K^*$ this would lie on top of the  $\omega (1390)$
 resonance found in the analysis  by Donnachie and Clegg\cite{donn},
which generally is assumed to be the radially excited $\omega$.
But, because of the P wave centrifugal barrier one cannot expect $1^{--}$
bound states from the potential (\ref{Vpv1}) alone without additional
short range attraction.

For the axial vector states $1^{+\pm}$, Eq.~(\ref{Vpvax}),
there are two channels like for the deuteron, and
we define  basis vectors  such that in general
$|1^{+\pm}>=u(r) |^3S_1>+w(r) |^3D_1>$.
The second  piece in Eq.~(\ref{Vpv1}) is given by the tensor potential
which connects the S and D waves just as for the deuteron.
The main difference compared to the deuteron is apart from the smaller
$\gamma_{IC}$, that the matrix
elements for the tensor $S_{12}$ are different.
 The off-diagonal element is smaller, and in the
D wave $<\ ^3D_1|S_{12}|^3D_1>=+1$ has changed sign, making the
tensor potental  attractive also in the D wave.

The overall strength of the potential
is given by $V_0\approx 1.3$ MeV and
the  relative coupling number $\gamma_{I\pm}$,
 whose absolute value is 3 for the isosinglet and 1 for the  isotriplet states.
The potential is positive  i.e. gives an attraction
for the  configurations listed in Table 3.
\vskip 0.5cm

$\begin{array}{lllll}
|\gamma_{I\pm}| & I &J^{PC}& P\bar V^*\ {\rm states}&
{\rm possible}\ K\bar K^*\ {\rm candidates}\\
\ \ 3     & 0 &0^{-+},1^{++},(1^{--}),... &
K\bar K^*,\  D\bar D^*, \  B\bar B^* & \eta(1440), f_1(1420),
(\omega(1390)?)\\
\ \ 1     & 1 &0^{--},1^{+-},(1^{-+}),... &
K\bar K^*,\ D\bar D^*,\ B\bar B^*  & (\hat\rho(1405)?)  \ .
\end{array}$
\vskip 0.3cm
\noindent{\small Table 3. The attractive $P\bar V$ channels with lowest
$J^{PC}$
and possible experimental $K\bar K^*$ candidates.}
\vskip .5cm

Observe that there are even two CP-exotic $0^{--},1^{-+}$ configurations,
but, the  number $|\gamma_{I\pm}|$ is only  $=1$,
which hardly is sufficient for producing bound states, but possibly it
can enhance the partial wave at threshold.
There is  evidence for this in $1^{-+}$
in the 1400 MeV region of the $\eta\pi$ channel ($"\hat\rho (1405)"$)
 (See Refs.\cite{PDG}, \cite {alde}).  It may be possible that
the $K\bar K^*$ threshold effect predicted this way through unitarity is
seen in the $\eta \pi$ channel.

The isosinglet $(P\bar V^*)_+$   pseudoscalar and axial channels
have the largest attraction, for which the well depth parametes are
\begin{eqnarray}
 s_{0^{-+}}&=& 0.5953\cdot 9MV_0/m_\pi^2=0.38\cdot (M/GeV)\ ,\label{welldps}\\
 s_{1^{++}}&=& 0.5953\cdot 3MV_0/ m_\pi^2=0.13\cdot (M/GeV)\ .\label{welldax}
\end{eqnarray}

These of course only take into account the Yukawa part, $e^{\mu r}/r$,
of the potentials (and for the same reason no well depth parameter exists
for the $1^-$ channel, since the Yukawa term is absent).
In the pseudoscalar case there is a
Yukawa term also from  the tensor potential
making the effective $ \gamma$ really 3 times bigger or 9
in  Eq.~(\ref{welldps}). This is even
slightly bigger than for the deuteron where $\gamma =8.33$.
 For  $B\bar B^*$ $s_{0^{+-}} \approx 2$ and $s_{1^{++}}\approx 0.68$, implying
that the pseudoscalar  must exist and that the axial state is extremely likely.

In the light meson sector  pseudoscalar a $(K\bar K^*)_+$ state
is a clear borderline case since $s_{0^-}$
is smaller by a factor of $(M_K+M_{K^*})/(M_D+M_{D^*})=2.8$,
making $s_{0^{-+}}= 0.26$ (and $s_{1^{++}}=0.09$ for the $f_1(1420)$ candidate)
or  smaller than for the deuteron.
Here the uncertainty of the nonrelativistic
approximations made (neglect of recoil, neglect of on shell pions etc.)
make definite conclusions difficult.
Certainly, with some additional attraction from non-pion exchanges, one has
the attractive possibility that  the lightest of the
iota peaks $\eta(1390-1490)$ and
the $f_1(1420)$ (c.f. Longacres paper\cite{long})
are $(K\bar K^*)_+$ deusons.

An especially interesting new situation appears
in the pseudoscalar channel as is
seen explicitely from Eq.~(\ref{Vpv0}): the tensor part with its $r^{-2}$
and $r^{-3}$ terms contributes, with
much stronger attraction than the central part, directly to the $^3P_0$
wave, and not through a second order coupling of S and D waves as for
the deuteron. This is  similar to  the $^3P_0$
NN case, Eq.~(\ref{VNN11}),  apart from the overall sign.

This means that
the $ r^{-2}$ and the $r^{-3}$  terms of the tensor potential
can compensate, at least partially, the centrifugal barrier $2/(Mr^2)$.
In Fig.~3  the potential (\ref{Vpv0}) is shown in the pseudoscalar, I=0
channel and the effective potential when the centrifugal barrier is added.
Although the result depends  on how the potential is regularized
at small $r$, the centrifugal barrier is
well compensated for the heavy mesons (see Fig.~3).
Also for $(K\bar K^*)_+$ partial compensation occurs, and
one can use this to argue that the centrifugal barrier is
no obstacle for interpreting  the iota peak as a P wave
 $(K\bar K^*)_+$ deuson.

Numerically by solving the Schr\"odinger equation, along the lines discussed
in Sec.~4, one finds in the
pseudoscalar channel that the one-pion
potential (\ref{Vpv0}) alone, regularized with $\Lambda =1.2$ GeV,
gives a bound state in $B\bar B^*$ with binding energy 58 MeV, i.e. an
$\eta_b (10545)$. In $D\bar D^*$ similarily one finds
an almost bound state near  threshold,  $\eta_c(\approx 3870)$,
(is bound if $\Lambda > 1.5$ GeV).
 In the light sector for $K\bar K^*$, one can get a
bound state or resonance (the $\eta (1440)?$)
if one adds a short range potential,
$\approx (-0.6{\rm GeV}) \cdot exp[-(r/0.9{\rm fm})^2]$, similar to those
used in Refs.\cite{WeinIsgur1},\cite{Barnes},\cite{Swanson}.

In the axial channel one finds numerically that for $B^*\bar B^*$
(with $\Lambda=1.2$ GeV, see Fig.~4)
 that there is a bound state with 41 MeV binding energy,
i.e. a $\chi_{b1}(10562)$, and in $D\bar D^*$ a state close to threshold,
$\chi_{c1}(\approx 3870)$. In the light sector in $K\bar K^*$ one can get a
bound state near threshold (the $f_1(1420)$?), if one
adds a rather
weak short range attraction $(-0.1{\rm GeV})\cdot exp[-(r/0.9{\rm fm})^2$.

To summarize I find
\begin{itemize}
\item that there certainly must exist
an $\eta_b(\approx 10545)$, that there very likely exists an $\eta_c(\approx
3870)$, and that possibly in the iota peak, $\eta(1440)$, there is a
$K\bar K^*$ deuson.
\item that certainly there is a $\chi_{1b}(\approx 10562)$ (Fig.~4),
that rather likely there must exist $\chi_{1c}(\approx 3870)$ and that
possibly the $f_1(1420)$ could be a $K\bar K^*$ deuson.
\end{itemize}

Provided, the heavy systems are not too deeply bound, such that annihilation
to light quarks could be large, the heavy  states should be narrow since
the decay to $D\bar D$ or $B\bar B$ is forbidden by parity.
\vskip 1cm

\noindent {\bf 5.2 Flavour exotic $PV$ states.}
\vskip 0.4cm

The Born term diagrams for flavour exotic $PV$
states are obtained from those discussed above
by line-reversing the $P$ and $V$ lines at the $PV\pi$ vertex.
There is then an overall  change of  sign because of  the pion G-parity.
One thus  has  $\gamma_I=-3$  for I=0 and +1 for I=1, which means that
only the latter gives a weakly
attractive potential for isovector $0^-$ and $1^+$ states, with
potentials of the same forms as given in Eqs.~(\ref{Vpv0}-\ref{Vpvax}).
For  $1^-$  states the dominant tensor part makes the overall
sign change as in Eq.~(\ref{Vpv1}). Therefore, one  has a relatively
large attraction ($|\gamma_0|=3$) in the isoscalar $1^-$ channels.
Summarizing the quantum numbers for which pion exchange gives some
attraction are listed in Table 4.
\vskip 0.5cm

$\begin{array}{llll}
|\gamma_{I}| & I &J^{P}            & {\rm States}\\
\ 1     & 1 &0^-, \ 1^+      &
KK^*,\ DD^*,\ BB^*,\ KD^*+K^*D,\ K\bar B^*+K^*\bar B,\ D\bar B^*+ D^*\bar B\
,\cr
\ 3     & 0 &1^-      &
KK^*,\ DD^*,\ BB^*,\ KD^*+K^*D,\ K\bar B^*+K^*\bar B,\ D\bar B^*+ D^*\bar B \ .
\end{array}$
\vskip 0.3cm

\noindent
{\small Table 4. As in Table 3, but for the flavour exotic $PV$ states.}
\vskip 0.5cm

The perhaps most likely place to find flavour exotic states would be in
isoscalar $BB^*$ with $J^P=1^-$, although the P wave centrifugal barrier
is an obstacle.
Should such states exist, which would be notoriously difficult
to produce, they must be very narrow, since
annihilation to light quarks is not possible. Pseudoscalars and axial
states could neither decay, because of parity,
to $BB$, implying that such states would be
essentially as stable as the $B$ and the $B^*$.
\eject
\onehead{\bf 6. VECTOR-VECTOR STATES.}

The vector-vector ($V\bar V$ AND $VV$)
 systems are the next lightest system where pion
exchange can be important in forming bound states.
\vskip 0.7cm

\noindent {\bf 6.1 Flavour nonexotic $V\bar V$ (and $V\bar V'$) states.}
\vskip 0.4cm

For composites of two vector mesons $V\bar V$ the relative coupling numbers
$\gamma^{VV}_{SI}$  are in the different spin isospin states given in
Table 5 below.
\vskip 0.5cm

$$\vbox {\halign {
# \hfill      & #\hfill&   \hfill#\hfill                 &  \hfill  #       \cr
            S & I\ \   &      States                     &$\gamma^{VV}_{SI}$\cr
            0 & 0\ \   & $^1S_0,\ ^1P_1,\ ^1D_2\ ...\ \ $&$       +6$       \cr
            1 & 0\ \   & $^3S_1,\ ^3P_J,\ ^3D_J\ ...\ \ $&$       +3$       \cr
            2 & 0\ \   & $^5S_2,\ ^5P_J,\ ^5D_J\ ...\ \ $&$       -3$       \cr
            0 & 1\ \   & $^1S_0,\ ^1P_1,\ ^1D_2\ ...\ \ $&$       -2$       \cr
            1 & 1\ \   & $^3S_1,\ ^3P_J,\ ^3D_J\ ...\ \ $&$       -1$       \cr
            2 & 1\ \   & $^5S_2,\ ^5P_J,\ ^5D_J\ ...\ \ $&$       +1$       \cr
\cr
}}$$
\noindent {\small Table 5.
The quantum numbers, partial waves and
value of the relative coupling number $\gamma^{VV}_{SI}$ for isodoublet
vector constituents.}
\vskip 0.8cm

As can be seen
the strongest attraction, as measured by $\gamma^{VV}_{SI}$,
is in the spin and isospin singlet channels ($\gamma^{VV}_{00}=+6$),
followed by the spin triplet, isosinglet channels ($\gamma^{VV}_{10}=+3$).
For I=1 one has either repulsion or very weak attraction.

$$\vbox {\halign {
\hfill # & # \hfill & \hfill # \hfill & \hfill # \hfill & \hfill # \hfill \cr
$J^{PC}$ & \ \ States  & $<\ |S_{12}|\ >$&
$\gamma^{VV}_{S0}$&$\gamma^{VV}_{S1}$   \cr
$0^{++}$&$\ \ ^1S_0\  ^5D_0 $& See Eq.~(\ref{Vvscalar})\ \ &$+6$&$-2$ \cr
$0^{-+}$&$\ \ ^3P_0         $& $+2$      &$+3$&$-1$   \cr
$1^{++}$&$\ \ ^5D_1$         & $-1$      &$-3$&$+1$   \cr
CP-exotic: $1^{-+}$&$\ \ ^3P_1 $& $-1$      &$+3$&$-1$   \cr
$1^{+-}$&$\ \ ^3S_1,\ ^3D_1 $& See Eq.~(\ref{Vv1pm})\ \  &$+3$&$-1$ \cr
$1^{--}$&$\ \ ^1P_1,\ ^5P_1,^5F_1$&See Eq.~(\ref{Vvector})\ \   &$+6$&$-2$ \cr
$2^{++}$&$\ \ ^1D_2,\ ^5S_2,\ ^5D_2,\ ^5G_2$\ \ \
&See Eq.~(\ref{Vvtensor})\ \  &$+6$&$-2$ \cr
CP-exotic: $2^{+-}$&$\ \  ^3D_2$&$-1$  &$+3$&$-1$\cr
}}$$

\noindent {\small Table 6. The lowest spin parity $VV$ states $J^{PC}$,
the contributing
spin orbit states, the relative magnitude of the tensor term,
 and the relative coupling numbers for I=0 and 1.}
\vskip 0.5cm

\noindent

 Table 6 lists the lowest spin parity states, $J^{PC}$,
and the contributing  partial waves.
There are also listed the relative coupling number
for isodoublet constituents  ($K^*\bar K^*,\ D^*\bar D^*,\ B^*\bar B^*$)
which is given by
$\gamma^{VV}_{SI} = -[(S+1)S-4)]\cdot [(I+1)I-3/2]$.
{}From Table 6 one sees that for
S waves, the largest $\gamma$'s appear  in the isoscalar $0^{++}$ channel,
for which $\gamma^{VV}_{00}=+6$, and in
the isoscalar, $1^{+-}$ channel, where
$\gamma^{VV}_{10}=+3$. In both of these channels
the S wave can mix with a D wave: $^5D_0$,
respectively $^3D_1$. Thus these are two-channel problems with the
potentials:

\begin{eqnarray}
V_{0^{++}}& =& -\gamma^{VV}_{0I} V_0   \left[
\left( \begin{array}{cc} 1&   0      \\ 0 & -\frac 1 2\end{array}\right) C(r) +
\left(\begin{array}{cc}
0&\sqrt\frac 1 2\\ \sqrt\frac 1 2&1\end{array}\right) T(r)
\right]\ , \label{Vvscalar}\\
V_{1^{+-}}& =& -\gamma^{VV}_{1I} V_0   \left[
\left( \begin{array}{cc} 1&   0      \\    0     & 1\end{array}\right) C(r) +
\left( \begin{array}{cc} 0& -\sqrt 2 \\ -\sqrt 2 & 1\end{array}\right) T(r)
\right]\ . \label{Vv1pm}
\end{eqnarray}

Here as in the following multichannel cases the overall $\gamma^{VV}_{SI}$ is
normalized by the first state which has the strongest central potential.
Note that, contrary to the case for the deuteron,  the tensor potential
in the D wave has the same sign as the scalar potential is in the S wave,
and that the coupling between the two waves is weaker.
Also, the central potential is in the scalar case, Eq.~(\ref{Vvscalar}),
 of opposite sign in the two waves.
Since the   $\gamma^{VV}_{SI}$ is larger by a factor of 2
in the scalar channel compared to the
axial channel scalar bound states are more likely. For the axial
$1^{+-}$ channel the potential  (\ref{Vv1pm})
 is identical to that of the $PV$  $1^{++}$
channel, Eq.~(\ref{Vpvax}).

For the single channel cases the most interesting
is the pseudoscalar $^3P_0$  where the tensor contributes
through the matrix element of $S_{12}=+2$.
This potential is given below together with the
three other single channel potentials:

\begin{eqnarray}
V_{0^{-+}} &=&
- \gamma^{VV}_{1I} V_0[C(r) + 2T(r)] = -\gamma^{VV}_{1I} V_0\frac {e^{-m_\pi
r}}
{fr} [3+\frac 6{m_\pi r}+\frac 6{(m_\pi r)^2}] \ ,\ \label{Vvps} \\
V_{1^{++}} =V_{1^{-+}} =V_{2^{+-}}&=&
-\gamma^{VV}_{SI} V_0[C(r) - T(r)]\  = +\gamma^{VV}_{SI} V_0
\frac {e^{-m_\pi r}}{fr}
[\frac 3{m_\pi r}+\frac 3{(m_\pi r)^2}] \ .  \label{Vv1pp}
\end{eqnarray}

The potential for the pseudoscalars, Eq.~(\ref{Vvps}), is the same
as  that for pseudoscalar $(P\bar V)_+$ states, (Eq.~\ref{Vpv0}),
and is  remarkably strong, as was already discussed for the $P\bar V$ case.
Therefore, the well depth parameter has the
same form as given in Eq.~(\ref{welldps})
of the previous section, and also here the
centrifugal P wave barrier is compensated by the
$r^{-2}$ and $r^{-3}$ terms in the tensor force.
Thus  one should expect pseudoscalar
bound states at least for  $B^*\bar B^*$ and probably also for $D^*\bar D^*$.
The $B^*\bar B^*$ states would be expected
quite near in mass with the $B\bar B^*$
states (roughly with same mass splitting as $M(B^*)-M(B)=46$ MeV).
Therefore, if the binding is of the same order of magnitude,
 one would in addition have to take into
account the mixing of these two states, since
pion exchange allows for the transition $B^*\bar B^*\to B\bar B^*$.
This shifts furthermore the lighter $B\bar B^*$ state down, making it stronger
bound.

For the channels $ 1^{++}=\ ^5D_1$,  $ 1^{-+}=\ ^3P_1$
and $ 2^{+-}=\ ^3D_2$ (the last two of which which are  CP-exotic)
the matrix element of $S_{12}$ is $-1$. Therefore,
one gets the weaker potentials of Eq.~(\ref{Vv1pp}) with opposite
overall sign, since
the central potential is cancelled by the first term
of the tensor potential. This is quite similar to the situation
in connection with the potential (\ref{Vpv1}) for the $P\bar V$ case.

Finally there remains two spin parities which  are of interest:
$1^{--}$ and $2^{++}$.
For the $1^{--}$, which is interesting since these states can be
produced directly in $e^+e^-$ annihilation, there are 3 channels,
 $(^1P_1,\ ^5P_1,\ ^5F_1)$ and the potential is:
\begin{eqnarray}
V_{1^{--}}= -\gamma^{VV}_{1I} V_0   \left[
\left(\begin{array}{cccc}1& 0  & 0    \\
                       0&-\frac 1 2& 0   \\
                       0& 0  &-\frac 1 2 \end{array}\right) C(r) +
\left(\begin{array}{cccc}
0                & +\sqrt\frac 4 5 &-\sqrt\frac 6 5     \\
+\sqrt\frac 4 5&  -\frac 7 5       &\sqrt\frac 6{25}    \\
-\sqrt\frac 6 5 &\sqrt\frac 6{25} &-\frac 8{5}          \\
\end{array}\right) T(r)
\right]\ .  \label{Vvector}
\end{eqnarray}

The overall $\gamma^{VV}_{10}=6$ is large in the isoscalar channels, but
the presence of at least a P wave centrifugal barrier in all partial waves
makes  bound states unlikely.

Finally for the $2^{++}$, there are 4 channels
$(^1D_2,\ ^5S_2, \ ^5D_2$,\ $^5G_2$), i.e. there is even an S wave present,
although it is repulsive in the isoscalar channel.
 The potential is:

\begin{eqnarray}
V_{2^{++}} = -\gamma^{VV}_{2I} V_0   \left[
\left(\begin{array}{cccc}1& 0  & 0  & 0  \\
                       0&-\frac 1 2& 0  & 0  \\
                       0& 0  &-\frac 1 2& 0  \\
                       0& 0  & 0  &-\frac 1 2\end{array}\right) C(r) +
\left(\begin{array}{cccc}
0                & +\sqrt\frac 1{10}&-\sqrt\frac 1 7  &+\sqrt\frac 9{35}   \\
+\sqrt\frac 1{10}&       0          &-\sqrt\frac 7{10}& 0                  \\
-\sqrt\frac 1 7 &-\sqrt\frac 7{10} &-\frac 3{14}     &-\frac{6\sqrt 5}{35}\\
+\sqrt\frac 9{35}& 0     &-\frac{6\sqrt 5}{35}&+\frac 5 7\end{array}\right)
T(r)
\right]\ . \label{Vvtensor}
\end{eqnarray}

 It seems difficult to
argue from this form, without numerical calculations, whether or not there can
be bound
states. As discussed below, quite surprisingly, this potential supports
bound states almost as deeply bound as the scalar and
pseudoscalar potential.

Numerically by solving the Schr\"odinger equation one finds that in the
pseudoscalar channel with the potential (\ref{Vvps}) and $\Lambda$=1.2 GeV
there is
a $B^*\bar B^*$ bound state with binding energy 59 MeV, i.e. an
$\eta_b(10590)$.
In  $D^*\bar D^*$ there is
almost a bound state at theshold (which becomes bound if
$\Lambda$ is increased to 1.5 GeV, or if one adds a very weak potential
such as $-0.05 [{\rm GeV}]\cdot exp[-(r/0.9{\rm [fm]})^2]$. In the light sector
$K^*\bar K^*$ becomes bound if one adds a potential
 $-0.42 [{\rm GeV}]\cdot exp[-(r/0.9{\rm [fm]})^2]$ giving roughly half of the
binding.

In the scalar sector the potential (\ref{Vvscalar}) with $\Lambda$=1.2 GeV
binds a $B^*\bar B^*$ $\chi_{b0}(10582)$ (see Fig.~5)
with a 67 MeV binding energy,
and a $D^*\bar D^*$ $\chi_{c0}(\approx 4015)$ near threshold (with 1 MeV
binding
energy). Again for the  $K^*\bar K^*$ system one must add a weak
potential to the central piece such as
$-0.1[{\rm GeV}]\cdot exp[-(r/0.9{\rm [fm]})^2]$ in order to have a bound
state.

In the axial sector, $1^{+-}$, the potential (\ref{Vv1pp}) supports a bound
state in $B^*\bar B^*$, i.e. a $h_b(10608)$ (see Fig.~4)
with a 41 MeV binding energy, and almost a
  $D^*\bar D^*$ bound state, a  $h_c(\approx 4015)$, at threshold (which
becomes bound if $\Lambda$ is increased to 1.3 GeV). For  $K^*\bar K^*$
one would have to add to the central piece some short range attraction,
such as $-0.2[{\rm GeV}]\cdot exp[-(r/0.9{\rm [fm]})^2]$ to have a bound state.

Finally in the 4-dimensional tensor case, Eq.~(\ref{Vvtensor}),
which is difficult to
analyse analytically, and where the S wave central potential is
repulsive for I=0, one numerically finds perhaps surprisingly
that one also gets a bound state in $B^*\bar B^*$ a $\chi_{b2}(10602)$
with 47 MeV binding (see Fig.~6.)
Furthermore one has an almost bound $D^*\bar D^*$
near threshold.
The reason for this is that, similar to what was
 the case for the deuteron, the  tensor force results in a mixing
between the partial waves,  such that an attraction is fed into the
S wave from the 3 higher partial waves through  higher order
iterative processes. In this solution the S wave has the highest
probability, 0.666, followed by the $^5D_2,^1D_2$ and $^5G_2$ wave
probabilities, 0.214, 0.106 and 0.014 respectively.

Summarizing from the above discussion one finds that one  can expect
$V\bar V$ bound deusons of the following kinds:
\vskip 0.5cm

\begin{itemize}
\item In the pseudoscalar sector an
 $\eta_b(\approx 10590)$ should exist and $\eta_c(\approx 4015)$
should very likely also exist
and an $ \eta (\approx 1790)$ is possibly bound.
These states should all mix to some extent with their $P\bar V$ counterparts,
making the lighter (mostly $P\bar V$) more bound.

\item In the scalar sector there should exist  a
$\chi_{b0} (\approx 10582)$ $B^*\bar B^*$ bound state (Fig.~5). A
$\chi_{c0}(\approx 4015)$ also likely exists,
and possibly the $f_0(1720)$ (the "theta") is a $K^*\bar K^*$ deuson.

\item In the axial sector
$h_b(\approx 10608)$ should exist (Fig.~4), possibly also a
$h_c(\approx 4015)$, while $ h_1(\approx 1790)$
is less likely.

\item In the tensor sector the numerical calculations show, perhaps
surprisingly, that there should exist a
$\chi_{b2} (\approx 10602)$ $B^*\bar B^*$ bound state (Fig.~6), and that
a $\chi_{c2}(\approx 4015)$ at threshold also is
very likely, while with some extra
attraction possibly a $f_2(\approx 1790)$ (the "theta" with spin 2?)
could exist.
\end{itemize}
\vskip 0.6cm

On the other hand one should
not expect axial ($1^{++}$) or vector ($1^{--}$) $V\bar V$ states
nor CP-exotic
($1^{-+},\ 2^{+-}$) states to be bound by the potential of Eq.~(\ref{Vv1pp}).
That  CP exotic states are not expected is, of course, unfortunate
from the point of view of not having an easily testable prediction.
But, since no such states have been observed these
negative results agree with the deuson model expectations.

Above the $V\bar V'$ states ($D^* \bar K^*,\ B^*\bar D^*,
...$) states have not been discussed,
but obviously they can be treated in a similar way.
E.g. $B^*\bar D^*$ states should certainly exist, but would be
more difficult to produce in the laboratory, sice they would have to be
produced together with another $D^*\bar B^*$ system.
\vskip 0.7cm

\noindent {\bf 6.2
Flavour exotic $VV$ and $VV'$ states, $K^*K^*,\ D^*D^*,\ B^*B^*$ and
$K^*B^*$ etc.}
\vskip 0.4cm

The above formulas can easily be generalized to the flavour exotic
composites, which are obtained by line-reversal of  the $V$-mesons
at one $VV\pi$ vertex. By the negative G-parity of the pion the over all
$\gamma^{VV}_{SI}$ changes sign,
and in addition one must take into the account the
Bose statistics when one has identical constituents.
Otherwise the same formulas (\ref{Vvscalar}-\ref{Vvtensor}) apply.
 Therefore, the isospin
(denoted in Table 7 by $I_1$)
will be fixed by the spin-parity of the state, whereby the overall
relative coupling number
$\gamma^{VV}_{SI_1}$, defined for the state with the largest attraction,
is also fixed and given in Table 7.

$$ \vbox {\halign
{ \hfill#&#&\hfill#\hfill&\hfill#\hfill&#\hfill&\hfill#\hfill&#&\hfill#\hfill&
\hfill#\hfill\cr
$J^{P}$& &$\ \ I_1\ \ $&$\ \ I_2\ \ $&\ States
& $<\ |S_{12}|\ >$ & &
$\ \gamma^{VV}_{SI_1}\ $ &$\ \gamma^{VV}_{SI_2}\ $ \cr
$0^{+}$&&1&0&$   ^1S_0,\ ^5D_0 $& See Eq.~(\ref{Vvscalar})
\ \ & &$+2$&$+3$\cr
$0^{-}$& &1&0&$   ^3P_0         $& +2       & & $+1$&$+3$\cr
$1^{+}$& &1&0&$   ^5D_1$         & $-1$     & & $-1$&$-3$\cr
$1^{-}$& &1&0&$^3P_1 $& $-1$     & &$+1$&$+3$\cr
$1^{+}$& &0&1&$   ^3S_1,\ ^3D_1$ & See Eq.~(\ref{Vv1pm})\ \ & &$-3$&$+1$\cr
$1^{-}$& &0&1&$   ^1P_1,\ ^5P_1,^5F_1$&See Eq.~(\ref{Vvector})
\ \    & &$+3$&$+2$\cr
$2^{+}$& &1&0&$   ^1D_2,\ ^5S_2,\ ^5D_2,\ ^5G_2$\ \ \
&See Eq.~(\ref{Vvtensor})\ \ & &$+2$&$+3$\cr
$2^{+}$& &0&1&$   ^3D_2 $& $-1$& &$-3$&$+1$\cr
}}$$
\vskip 0.3cm

{\small Table 7. As in Table 5 but for the flavour exotic $VV$ states.
The isospin for identical mesons is given by $I_1$,
while $I_2$ is allowed only when the mesons are nonidentical. The last two
colums show the largest values of $\gamma^{VV}_{SI_1}$ and
$\gamma^{VV}_{SI_2}$.  }
\vskip 0.5cm

{}From the values of $\gamma^{VV}_{SI_1}$ one sees that one never has as large
attraction
as in the flavour nonexotic $V\bar V$ case.
The strongest attraction measured by the
largest  $\gamma^{VV}_{SI_1}$ appear  for the 3-dimensional  $1^-$ state,
where however, centrifugal barriers exist.
The next strongest attraction
($\gamma_{SI}=+2$) are in
scalar, and tensor I=1 $B^{*}B^{*}$ states.

When the vector mesons are different
as in $K^*B^*$, $K^*\bar D^*$ or $\bar D^* B^*$ Bose statistics does not
put restriction on the isospin and can have the values $I_2$  of  Table 7.
 But, also  there  $\gamma^{VV}_{SI_2}$ is never very large
 as seen from Table 7.
In  the $^5D_1$ I=0 channel, where $\gamma^{VV}_{01}$
is ($-3$) the central and tensor parts cancel  in Eq.~(\ref{Vv1pp}).
Also a relatively large  $\gamma^{VV}_{SI_2}$ is
in the 4-dimensional $2^+$ channel with I=1, where in the S wave
$\gamma^{VV}_{20}=+3$.

This means that flavour exotic $VV$ as well as
$VV'$ bound states are unlikely except possibly when the constituents
are the heavy $B^*$:s
\vskip 2cm

\onehead{\bf 7. CONCLUSIONS}
In this paper the question whether deuteronlike meson-meson
bound states or deusons exist have been analysed,
and it was found that in particular in the
heavy meson sector such states certainly must exist. In Table 8 the
quantum numbers of altogether 12 expected heavy
deusons are listed. Especially in
the beauty sector the existence of these states seems impossible to avoid,
 since they are already bound
by about 50 MeV including only one-pion exchange.  In the charm sector
pion exchange alone predicts states near the thresholds, and with some small
contribution of shorter range these states should also exist.

$$\vbox {\halign {\hfill
# \hfill      &  #\hfill &   \hfill#                            & \hfill #\cr
   Composite  & $J^{PC}$\ \ \ &     Deuson\ \ \ \               \cr
  $D\bar D^*$ & $0^{-+}$& $\eta_c    (\approx 3870) $          \cr
  $D\bar D^*$ & $1^{++}$& $\chi_{c1} (\approx 3870) $                   \cr
$D^*\bar D^*$ & $0^{++}$& $\chi_{c0} (\approx 4015) $                   \cr
$D^*\bar D^*$ & $0^{-+}$& $\eta_{c}  (\approx 4015) $                   \cr
$D^*\bar D^*$ & $1^{+-}$& $   h_{c0} (\approx 4015) $                   \cr
$D^*\bar D^*$ & $2^{++}$& $\chi_{c2} (\approx 4015) $                   \cr
  $B\bar B^*$ & $0^{-+}$& $\eta_b    (\approx 10545)$                   \cr
  $B\bar B^*$ & $1^{++}$& $\chi_{b1} (\approx 10562)$                   \cr
$B^*\bar B^*$ & $0^{++}$& $\chi_{b0} (\approx 10582)$                   \cr
$B^*\bar B^*$ & $0^{-+}$& $\eta_{b}  (\approx 10590)$                   \cr
$B^*\bar B^*$ & $1^{+-}$& $   h_{b} (\approx 10608)$                   \cr
$B^*\bar B^*$ & $2^{++}$& $\chi_{b2} (\approx 10602)$                   \cr
\cr
}}$$
\noindent {\small Table 8.
The predicted heavy deuson states
(all with I=0) close to the $D\bar D^*$ and the
$D^*\bar D^*$ thresholds and about 50 MeV below the $B\bar B^*$ and
$B^*\bar B^*$ thresholds. As discussed in the text, the mass values are
obtained from a rather conservative one-pion exchange contribution only.
With additional attraction of shorter range, the masses can decrease
considerably. Mixing between the two $\eta_b$'s (and two $\eta_c$'s)
should decrease the the lighter mass somewhat (and increase the heavier mass).
}
\vskip 0.8cm

The widths are expected to be quite narrow. This is especially the case
for the four  $D\bar D^*$ and $B\bar B^*$ states, which because of parity
cannot
decay to $D\bar D$, respectively $B\bar B$. Of course, through annihilation
of the heavy quarks, decays to light mesons are possible.
However, this should be
suppressed by form factors, since these states are
much larger in size than normal $Q\bar Q$ states,
provided they are weakly bound.
The $D^*\bar D^*$ and $B^*\bar B^*$ deusons can generally decay into
$D\bar D$ and $B\bar B$, which should be their main decay mode giving
widths of a few tens of MeV.

The lighter the constituents are, the larger will the kinetic term of the
Hamiltonian be compared to the potential term. Therefore, deusons with light
mesons as constituents  are much harder to bind,
and one-pion exchange alone cannot support such bound states. With the pion
itself as a constituent this is definitely a crucial obstacle, but also
for $K\bar K^*$ systems
one has a potential term roughly only half as strong as needed. Of course,
this conclusion depends somewhat on how the tensor potential is regularized.
(One could argue that the cut off $\Lambda$
should be larger, and  furthermore, one could increase the scale
$V_0=1.3$ MeV to 1.75 MeV, as suggested by the $K^*$ width,
Eq.~(\ref{width}). Both of  these arguments
would increase the binding.) But, it should also be remembered that the
nonrelativistic potential model is least reliable for
light hadrons, and that in  $K\bar K^*$ the intermediate pion can be on
shell, which should give some repulsion.

In Table 9 the quantum numbers are listed, where one finds
 the largest attraction
in the light meson sector, together with the experimental non-$q\bar q$
candidates. The fact that all the best non-$q\bar q$ candidates are
included in this list, can be taken as support of the idea that these
states actually are deusons, where pion exchange contribute about half
of their binding energy.

$$\vbox {\halign {\hfill
# \hfill &\hfill # &    # \hfill   &\hfill#\hfill  &  #\hfill \cr
Composite&\ \ \ $I$& $J^{PC}$\ \ \ & Threshold/MeV &\ \ \ Experimental
non-$q\bar q$
state\cr
  $K\bar K^*$ &0& $0^{-+}$& 1390 &$\ \ \ \eta(1390)?$ in the "iota" peak \cr
  $K\bar K^*$ &0& $1^{++}$& 1390 &$\ \ \ f_1 (1420) $        \cr
$K^*\bar K^*$ &0& $0^{++}$& 1790 &$\ \ \ f_0 (1710) $ "theta"\cr
$K^*\bar K^*$ &0& $0^{-+}$& 1790 &$\ \ \ \eta (1760)$\cr
$K^*\bar K^*$ &0& $1^{+-}$& 1790 &$          $\cr
$K^*\bar K^*$ &0& $2^{++}$& 1790 &$\ \ \ f_2(1720)$ "theta"\cr
$(\rho\rho +\omega\omega)/\sqrt 2$&0& $0^{-+}$& 1540-1566  &$\ \ \
\eta(1490)$? in the "iota" peak\cr
$(\rho\rho -\omega\omega)/\sqrt 2$&0& $0^{++}$& 1540-1566 &$\ \ \
f_0(1515)$\cr
$(\rho\rho +\omega\omega)/\sqrt 2$&0& $2^{++}$& 1540-1566 &$\ \ \
f_2(1520)$ ("Ax")\cr
$(K^*\rho -K^*\omega)/\sqrt 2$    &$\frac 12$& $0^{++}$& 1665-1678 & \cr
\cr
}}$$
\noindent {\small Table 9.
The meson-meson channels in the light meson sector
where one-pion exchange is attractive. With
additional attraction of
shorter range, giving roughly half of the attraction,
bound states should exist. The  experimental non-$q\bar q$
candidates which could be such  deusons are listed in the last column.
The iota peak (listed under $\eta (1440)$
in Ref.\cite{PDG}) probably  contains
two states $\eta (1390)$ and $\eta (1490)$ and the theta peak, $f_0(1710)$, may
contain both spin 0 and 2 (See Ref.\cite{PDG},\cite{theta}).
The $(\rho\rho\pm\omega\omega)/\sqrt 2$ and $(K^*\rho -K^*\omega)/\sqrt 2$
channels were mentioned in Ref.\cite{NAT1},
and are not discussed further in this paper,
being left for future work.}

\vskip 0.9cm

In channels with exotic flavour or CP quantum numbers
pion exchange is generally repulsive or
quite weak. Therefore, one does not expect
that such exotic deusons  exist, although
$B^* B^*$ may be an exception.
Neither does the deuson
model predict new non-$q\bar q$  states which should have been
seen. E.g., for I=1 channels
one pion exchange is generally
a factor 3 weaker than for I=0, and one  certainly does not expect such states
within the light meson sector. Also for $J^{PC}=1^{--}$ the centrifugal
barrier essentially forbids such deusons from
existing (which should have been seen in $e^+e^-$ annihilation).

Where could the predicted heavy deusons of Table 8 be produced and seen
experimentally? Unfortunately, this will not be easy, but
two good places seems to be:  $p\bar p$
in flight at the Fermilab antiproton accumulator, and $\Upsilon$ decay
looking at final states including e.g.
$J/\psi\ \omega $ (for the $D\bar D^*$ deusons)
or $D\bar D$, $D\bar D^*$ (for the $D^*\bar D^*$ deusons).

To find these states would be important, not only because they would
confirm the expectations from pion exchange and constrain the parameters
of the model presented here. More importantly, if these
deusons are found, they at the same time
would give strong support for the interpretation that
many, perhaps all, of the
present light non-$q\bar q$ candidates really are deuteronlike states.
 This would then
imply that experimental evidence for baglike multiquark states and glueballs
would have to be looked for at higher energies.
\bigskip

{\it Acknowledgements.} I thank J.-M. Richard  for discussions and
for pointing out the numerical method used in this paper. Also comments by
T.E.O. Ericson, A.M. Green, G. Karl and J. Paton
 have been useful in this work.
{\bf Figure captions.}
\vskip 0.7cm
\setcounter{page}{20}

Fig.~1. The central potential $C(r)$, Eq.~(\ref{Cr}),
(normalized as for the deuteron by a factor
11.2 MeV) and the tensor potential $T(r)$,
eq.~(\ref{Tr}) as function of the distance $r=r_1-r_2$ between the constituents
together with their regularized forms $C_\Lambda (r),\ T_\Lambda (r)$,
c.f. eqs. (\ref{Creg}-\ref{Treg}) with $\Lambda =1.2$ GeV.
\bigskip

Fig.~2. The potential for the I=0,
pseudoscalar $^3P_0$ $P\bar V$ and $V\bar V$
channels (solid line) and the effective potentials for
$K\bar K^*$, $K^*\bar K^*$,
$D\bar D^*$ and $B\bar B^*$ systems when the centrifugal barrier $2/(Mr^2)$
is added. As can be seen for $D\bar D^*$ and heavier constituents the
centrifugal barrier is overcome by the potential for distances greater than
0.5 fm.
\bigskip

Fig.~3. The deuteron $^3S_1$ and $^3D_1$ wavefunctions obtained using
the numerical method of Sec.~4 and a pure one-pion exchange potential
eq. (\ref{Vdeut}) with $\Lambda = 0.782$ GeV.
This solution agrees apart from a small
region near $r_1-r_2=0$ with solutions obtained using more detailed potentials
such as that of Ref.~\cite{Paris}.
\bigskip

Fig.~4. The $D^*\bar D^*$ deuson wave functions $^1S_0$ and $^5D_0$
for spin $0^{++}$ for a pure one-pion exchange potential, eq. (\ref{Vvscalar}),
assuming
$\Lambda$=1.2 GeV, and $V_0$=1.33 MeV. The solution resembles strongly
that of the deuteron. The binding energy for this solution is 1.2 MeV,
which  determines the exponential fall-off of the wave functions at large
distances. The probabilities of the different waves are
$P(^1S_0)=0.937,$ $P(^5D_0)=0.063$.
\bigskip

Fig.~5. As in Fig.~4 but for $B\bar B^*$ or $B^*\bar B^*$
axial deuson wave functions using the potential of eqs.~(\ref{Vpvax},
\ref{Vv1pm}). The binding energy is for this solution
41 MeV, and the probabilities of the different waves are
$P(^3S_1)=0.668,$ $P(^3D_1)=0.332$.
\bigskip

Fig.~6. As in Fig.~4 but for $B\bar B^*$
tensor, $2^{++}$, deuson wave functions using the
potential eq.~(\ref{Vvtensor}).
The binding energy is for this solution
47 MeV, and the probabilities of the different waves are
$P(^5S_2)=0.666,$ $P(^5D_2)=0.214,$ $ P(^1D_2)=0.106,$ $ P(^5G_2)=0.014$.
\eject
\vfill
\null
\vfill
\eject
\null
\eject
\null
\eject


\small

\begin{thebibliography}{99}
\setcounter{page}{24}
\bibitem{PDG} K. Hikasa et al., (The Particle Data Group),
 Phys. Rev. {\bf D45}  (1992) 1.

\bibitem{NAT1} N.A. T\"ornqvist,
Phys. Rev. Lett. \bf 67 \rm (1992) 556; \\
Proc. of Hadron 91, College Park, Maryland USA, 1991,
Eds. S. Oneda, D.C. Peaslee, World Scientific (1992),  795.

\bibitem{NAT2} N.A. T\"ornqvist, Proc. Int. Conf. High Energy Physics, Dallas
1992, Ed. J.R. Sanford, AIP (Conf Proc) No {\bf 272}, 784.

\bibitem{Okun} M.B. Voloshin and L.B. Okun, Pisma Zh.Exp.Teor.Fiz. \bf 23 \rm
(1976) 369, JETP Lett. \bf 23 \rm (1976) 333.

\bibitem{Masud} M. Chaichian, R. K\"ogerler, and M. Roos, Nucl. Phys. \bf B141
\rm (1978) 110.


\bibitem{Kramer} F. Gutbrod, G. Kramer and Ch. Rumpf, Z. Physik \bf C1 \rm
(1979) 391.

\bibitem{Barnes} K. Dooley, E.S. Swanson and T. Barnes,  Phys. Lett.
{bf B275} (1992)  478.

\bibitem{BarnesSw} T. Barnes and E.S. Swanson, Phys. Rev. \bf D46 \rm (1992)
 131.

\bibitem{Swanson} E.S. Swanson,
Ann.  Phys. (New York) {\bf 220} (1992) 73.

\bibitem{long} R. Longacre,  Phys. Rev. {\bf D42} (1990) 874 .

\bibitem{WeinIsgur1} J. Weinstein and N. Isgur, Phys. Rev. Lett.
\bf 48 \rm (1982)  656; Phys. Rev. \bf D27 \rm (1983) 588.

\bibitem{WeinIsgur2} J. Weinstein and N. Isgur, Phys. Rev. \bf D41 \rm (1990)
2236

\bibitem{WeinIsgur3} J. Weinstein and N. Isgur, Phys. Rev. \bf D43 \rm (1991)
95.

\bibitem{ISGUR} K. Maltman and N. Isgur, Phys. Rev. Lett. {\bf 50} (1983) 1827;
  Phys. Rev. {\bf D29} (1984) 952;\\ N. Isgur,  Acta Phys.
Austriaca  Suppl. {\bf XXVII} (1985) 177-266.

\bibitem{Karl} T.E.O. Ericson and G. Karl,
Physics Letters {\bf B 309} (1993) 426.

\bibitem{Mano}  A.V. Manohar and M.B. Wise,
 Nucl. Phys. {\bf B 399 } (1993) 17.

\bibitem{Glen} N.K. Glendenning and G. Kramer, \bf 126 \rm (1962) 2159.

\bibitem{Rosa} T.E.O. Ericson and M Rosa-Clot, Ann.Rev.Nucl.Part.Sci.
{\bf 35} (1985) 271, and Nucl. Phys. {\bf A405} (1983) 497.

\bibitem{Torleif} T.E.O. Ericson and M. Weise, Pions and Nuclei,
 Clarendron Press (1988).

\bibitem{Yuk} H.~Yukawa, Proc. Phys. Math. Soc. Japan {\bf 17} (1935) 48.

\bibitem{Pion}
C.M.G. Lattes, H. Muirhead, G.P.S. Occhialini, and C.F. Powell,
Nature {\bf 159} (1947) 594.

\bibitem{ACCMOR} S. Barlag et al., (ACCMOR Collaboration),
 Phys.Lett. {\bf B279} (1993) 480.

\bibitem{Thews} R.L. Thews and A.N. Kamal, Phys. Rev. {\bf D32} (1985) 810.

\bibitem{holinde} K. Holinde, Phys. Reports {\bf 68} (1981) 121.

\bibitem{BW} J.M. Blatt and V.F. Weisskopf, "Theoretical Nuclear Physics",
John Wiley \& Sons 1952, (page 56).

\bibitem{Basd} S. Boukraa and J.-L.Basdevant, J. Math. Phys. {\bf 30} (1989)
1060.

\bibitem{Rich} J.M  Richard, Phys. Rep. {\bf 212} (1992) 1.

\bibitem{Paris} M. Lacombe, B. Loiseau, J.M. Richard, R. Vinh Mau,
J. C\^ot\'e, P. Pir\`es and R. de Tourreil,  Phys. Rev. {\bf C21} (1990) 861.

\bibitem{donn} A. Donnachie and  A.B. Clegg, Z. Physik {\bf C42} (1989) 663;
ibid. {\bf C51} (1991) 689.

\bibitem{alde} D. Alde et al.. Phys. Lett. {\bf B205} (1988) 397.

\bibitem{theta}
In central production a $f_2(1720)$  resonance is seen with spin 2:
  A. Breakstone et al. Z.Physik {\bf C 58} (1993) 251. On the other hand
in $J/\psi$ decay spin 0 is favoured: L.P. Chen, SLAC-PUB 5669,
Proc. of Hadron 91, College Park, Maryland USA, 1991,
Eds. S. Oneda, D.C. Peaslee, World Scientific (1992), p. 111.

\end{thebibliography}
\end{document}